\documentclass[letterpaper]{jpsj-suppl}
\usepackage{txfonts} 

\title{Phase Transitions with Discrete Symmetry Breaking in Antiferromagnetic Heisenberg Models on a Triangular Lattice}

\author{
Ryo \textsc{Tamura}$^1$,
Shu \textsc{Tanaka}$^2$,
and
Naoki \textsc{Kawashima}$^3$
}

\inst{
$^1$International Center for Young Scientists, National Institute for Materials Science,
1-2-1, Sengen, Tsukuba-shi, Ibaraki, 305-0047, Japan\\
$^2$Department of Chemistry, University of Tokyo,
7-3-1, Hongo, Bunkyo-ku, Tokyo, 113-0033, Japan\\
$^3$Institute for Solid State Physics, University of Tokyo,
5-1-5, Kashiwanoha, Kashiwa-shi, Chiba, 277-8581, Japan
}

\email{tamura.ryo@nims.go.jp}

\recdate{}

\abst{
We study phase transition behavior of the Heisenberg model on a distorted triangular lattice with competing interactions.
The ground-state phase diagram indicates that underlying symmetry can be changed by tuning parameters.
We focus on two cases in which a phase transition with discrete symmetry breaking occurs.
The first is that the order parameter space is SO(3)$\times C_3$.
In this case, a first-order phase transition, with threefold symmetry breaking, occurs.
The second has the order parameter space SO(3)$\times Z_2$.
In this case, a second-order phase transition occurs with twofold symmetry breaking.
To investigate finite-temperature properties of these phase transitions from a microscopic viewpoint, we introduce a method to make the connection between continuous frustrated spin systems and the Potts model with invisible states.
}

\kword{frustration, phase transition, order parameter space, discrete lattice rotational symmetry, Potts model with invisible states}

\begin{document}
\maketitle

\section{Introduction}

Phase transitions in frustrated spin systems have been investigated for a long time\cite{Toulouse-1977,Liebmann-1986,Kawamura-1998,Diep-2005,Southern-2012}.
In frustrated spin systems, several types of order parameter spaces emerge, depending on the geometric structure of interactions and symmetry of spins. 
In many cases, the order parameter space can be categorized by the ground state.
In the Heisenberg spin systems, when the ground state is a non-collinear spin structure, the order parameter space is SO(3), whereas when the ground state is a collinear spin structure, the order parameter space is $S_2$.
In two-dimensional Heisenberg spin systems, the long-range order of spins is prohibited by the Mermin-Wagner theorem\cite{Mermin-1966}. 
However, the $Z_2$ vortex dissociation occurs when the order parameter space is SO(3)\cite{Kawamura-1984}. 
On the other hand, no phase transition occurs in systems where the order parameter space is $S_2$.

\begin{figure}[t]
\includegraphics[scale=1]{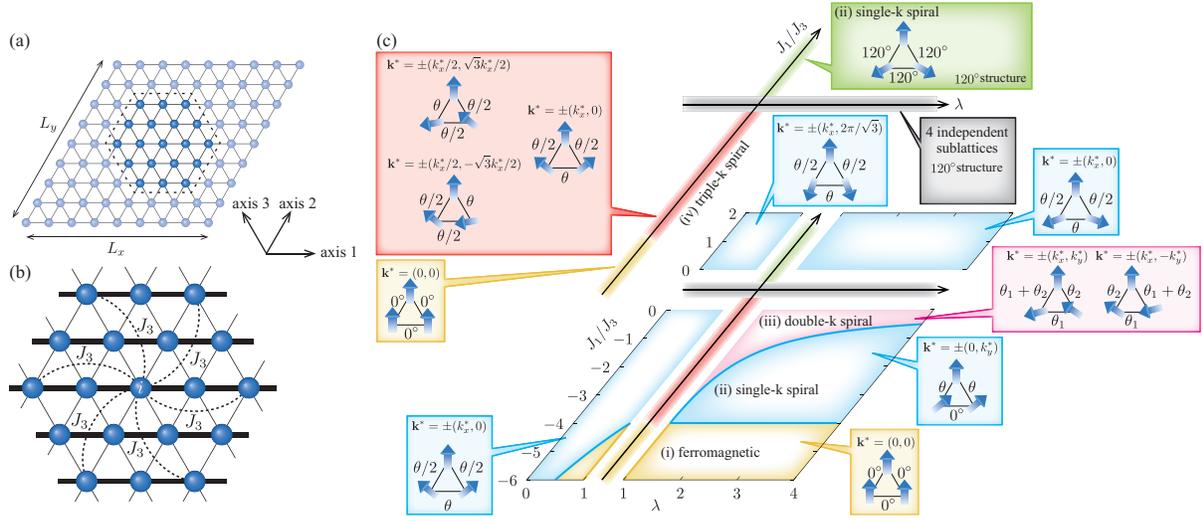}
\caption{
(a) Triangular lattice with $L_x \times L_y$ sites.
(b) Enlarged view of the dotted hexagonal area in (a). The thick and thin lines indicate $\lambda J_1$ and $J_1$, respectively.
The third nearest-neighbor interactions at the $i$-th site are depicted.
(c) Ground-state phase diagram of the model given by Eq.~(\ref{eq:Hamiltonian}).
Ground states can be categorized into five types.
More details in each ground state are given in the main text.
}
\label{fig:lattice}
\end{figure}

Recently, phase transitions in systems with the order parameter space described by the direct product between global rotational symmetry of spins and lattice rotational symmetry have been studied\cite{Chandra-1990,Loison-2000,Weber-2003,Tamura-2008,Stoudenmire-2009,Okumura-2010,Tamura-2011,Okubo-2012,Jin-2012,Tamura-2013,Tamura-2013b}.
In the Heisenberg model with the nearest-neighbor interaction $J_1$ and third nearest-neighbor interaction $J_3$ ($J_1$-$J_3$ model) on a triangular lattice, a first-order phase transition with threefold symmetry breaking occurs for a certain parameter region\cite{Tamura-2008,Tamura-2011}.
Phase transitions with discrete symmetry breaking can be often described by the ferromagnetic Potts model. 
For example, 3-state ferromagnetic Potts model in two dimensions exhibits a second-order phase transition with threefold symmetry breaking, which is not consistent with the abovementioned phase transition in the $J_1$-$J_3$ model on a triangular lattice. 
To overcome the contradiction, a generalized Potts model, called the Potts model with invisible states, was introduced\cite{Tamura-2010,Tanaka-2011a,Tanaka-2011b}. 
A first-order phase transition with threefold symmetry breaking occurs in the 3-state Potts model with invisible states.
In this paper, we consider a microscopic nature of the phase transition in the Heisenberg model with competing interactions on a distorted triangular lattice depicted in Figs.~\ref{fig:lattice}(a) and (b) from the viewpoint of the Potts model with invisible states.

\section{Model and Ground State Phase Diagram}

We consider the classical Heisenberg model on a distorted triangular lattice.
The Hamiltonian is given by
\begin{eqnarray}
 \label{eq:Hamiltonian}
 {\cal H} = \lambda J_1 \sum_{\langle i,j \rangle_{{\rm axis}\,1}} {\bf s}_i \cdot {\bf s}_j
  + J_1 \sum_{\langle i,j \rangle_{{\rm axis}\,2,3}} {\bf s}_i \cdot {\bf s}_j
  + J_3 \sum_{\langle\langle i,j \rangle\rangle} {\bf s}_i \cdot {\bf s}_j,
\end{eqnarray}
where the first term represents the nearest-neighbor interactions along axis 1, the second term denotes the nearest-neighbor interactions along axes 2 and 3, and the third term is the third nearest-neighbor interactions (see Figs.~\ref{fig:lattice}(a) and (b)).
The variable ${\bf s}_i$ is the three-dimensional vector spin of unit length. 
The parameter $\lambda (>0)$ represents a uniaxial distortion along axis 1.
Here we consider the case that the third nearest-neighbor interaction $J_3$ is antiferromagnetic ($J_3>0$).
The ground state of the model given by Eq.~(\ref{eq:Hamiltonian}) is represented by the wave vector ${\bf k}^*$ at which the Fourier transform of interactions $J({\bf k})$ is minimized.
In this case, $J({\bf k})$ is given by
\begin{eqnarray}
 \label{eq:groundstate}
  \frac{J({\bf k})}{N J_3} =
  \frac{\lambda J_1}{J_3}\cos k_x 
  + \frac{2J_1}{J_3}\cos\frac{k_x}{2}\cos\frac{\sqrt{3}k_y}{2}
  + \cos 2k_x + 2\cos k_x \cos\sqrt{3}k_y,
\end{eqnarray}
where $N(=L_x\times L_y)$ is the number of spins.
Here the lattice constant is set to unity.
It should be noted that the spin structures denoted by ${\bf k}$ and $-{\bf k}$ are the same in the Heisenberg model.
Figure~\ref{fig:lattice} (c) depicts the ground-state phase diagram, which shows five types of ground states, depending on the parameters.
First, let us consider the case of $J_1/J_3=0$ which is represented by the black area. 
In this case, the Hamiltonian given by Eq.~(\ref{eq:Hamiltonian}) is equivalent to four independent antiferromagnetic Heisenberg models on a triangular lattice, with only nearest-neighbor antiferromagnetic interaction. 
Then, the ground state is a $120^\circ$ structure. 
Next we consider the case of $J_1/J_3 \neq 0$. 
The ground state for $J_1/J_3 \neq 0$ can be categorized into four types in terms of the character of the wave vector ${\bf k}^*$.
In the ferromagnetic state indicated by (i), the order parameter space is $S_2$ and any finite-temperature phase transition is prohibited by the Mermin-Wagner theorem\cite{Mermin-1966}.
In the single-$k$ spiral state indicated by (ii), the order parameter space is SO(3), and the $Z_2$ vortex dissociation occurs.
When $J_1/J_3 > 0$ and $\lambda = 1$, the ground state is a $120^\circ$ structure which is also a single-$k$ spiral spin state.
In the double-$k$ spiral state indicated by (iii), the order parameter space is SO(3)$\times Z_2$.
A second-order phase transition with twofold symmetry breaking and the $Z_2$ vortex dissociation happen simultaneously\cite{Tamura-2013}.
In the triple-$k$ spiral state indicated by (iv), the order parameter space is SO(3)$\times C_3$.
A first-order phase transition with threefold symmetry breaking and the $Z_2$ vortex dissociation occur simultaneously\cite{Tamura-2008,Tamura-2011}.

\section{Phase Transition from the Viewpoint of the Potts Model with Invisible States}

\begin{figure}[b]
\includegraphics[scale=1]{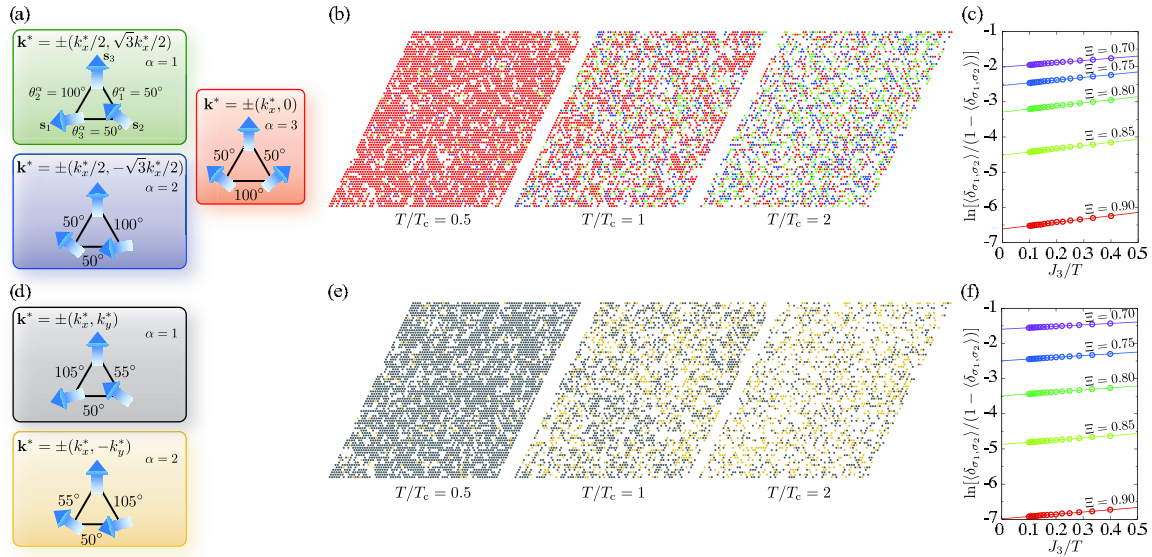}
\caption{
(a)-(c) SO(3)$\times C_3$ case.
(a) Three types of ground states.
(b) Snapshots at several temperatures on a $72\times 72$ triangular lattice.
(c) Analysis using Eq.~(\ref{eq:analysis}) with $q=3$ for several $\Xi$'s.
(d)-(f) SO(3)$\times Z_2$ case.
(d) Two types of ground states.
(e) Snapshots at several temperatures on a $72\times 72$ triangular lattice.
(f) Analysis using Eq.~(\ref{eq:analysis}) with $q=2$ for several $\Xi$'s.
}
\label{fig:snapshot}
\end{figure}

As shown in the previous section, a first-order phase transition with threefold symmetry breaking occurs for a certain parameter region, indicated by (iv) whereas a second-order phase transition with twofold symmetry breaking occurs for a different parameter region, indicated by (iii).
Both phase transitions also appear in the Potts model with invisible states\cite{Tamura-2010,Tanaka-2011a,Tanaka-2011b}.
Then, for correspondence between the model given by Eq.~(\ref{eq:Hamiltonian}) and the Potts model with invisible states, we introduce a locally defined parameter for each upward triangle, which quantifies the similarity between the present state and the ground state. 
First, we consider the case that the order parameter space is SO(3)$\times C_3$, in which a first-order phase transition with threefold symmetry breaking occurs.
The relative angles between spins on each triangle in the ground state configurations are denoted by $\{\theta_\ell^{\alpha}\}$, where $\ell$ labels the sides of each triangle, and $\alpha$ assigns the ground state.
In this case, since there are three types of ground states, $\alpha=1,2,3$ [see Fig.~\ref{fig:snapshot} (a)].
Let $\{{\bf s}_\ell\}$ be the spin variable on each triangle, as shown in Fig.~\ref{fig:snapshot} (a).
We define the following quantity $\xi^\alpha$ on each triangle by
\begin{eqnarray}
 \label{eq:localquantity}
 \xi^\alpha := 1 - \frac{1}{3}
\left[ \frac{ | \theta_1^\alpha-\cos^{-1} (\mathbf{s}_2 \cdot \mathbf{s}_3) |}{\max \{ |\pi-\theta_1^\alpha | , \theta_1^\alpha | \}}+
\frac{ | \theta_2^\alpha-\cos^{-1} (\mathbf{s}_3 \cdot \mathbf{s}_1) |}{\max \{ |\pi-\theta_2^\alpha | , \theta_2^\alpha | \}}+
\frac{ | \theta_3^\alpha-\cos^{-1} (\mathbf{s}_1 \cdot \mathbf{s}_2) |}{\max \{ |\pi-\theta_3^\alpha | , \theta_3^\alpha | \}} \right].
\end{eqnarray}
This quantity should be positive by definition.
When $\xi^\alpha=1$, the present state $\{{\bf s}_\ell\}$ is the same as the ground state indicated by $\alpha$.
As $\xi^\alpha$ becomes small, the difference between the present state and the ground state indicated by $\alpha$ becomes large.
The present state can be characterized by $\alpha'$ such that $\alpha'={\rm argmax}_\alpha \xi^\alpha$.
Since we now consider a phase transition with threefold symmetry breaking, three unit vectors are introduced: $\mathbf{e}_1 = (1,0)$, $\mathbf{e}_2= (-1/2,\sqrt{3}/2)$, and $\mathbf{e}_3=(-1/2,-\sqrt{3}/2)$ as the case of the 3-state ferromagnetic Potts model. 
A two-dimensional vector on each triangle is defined as $\boldsymbol{\eta}={\bf e}_{\alpha'} H(\xi^{\alpha'} - \Xi)$, where $H(\cdot)$ is the Heaviside step function and $\Xi$ represents the artificially introduced threshold, {\it i.e.} when $\xi^{\alpha'} < \Xi$, $\boldsymbol{\eta}=\boldsymbol{0}$.
Using the local parameter $\boldsymbol{\eta}$, the Heisenberg model given by Eq.~(\ref{eq:Hamiltonian}) connects to the Potts model with invisible states\cite{Tamura-2010,Tanaka-2011a,Tanaka-2011b}.
Here, $\boldsymbol{\eta}=\boldsymbol{0}$ is regarded as the invisible state.

We use the parameter sets $J_1/J_3=-0.73425$ and $\lambda =1$ in which the transition temperature is $T_{\rm c}/J_3=0.4746(1)$ and 
the relative angles $\{\theta_\ell^{\alpha}\}$ in the ground state are shown in Fig.~\ref{fig:snapshot} (a)\cite{Tamura-2011}.
Figure \ref{fig:snapshot} (b) shows snapshots of the local parameters in the case of $\Xi=0.8$ at several temperatures for $L_x=L_y=72$ with the periodic boundary conditions, which are obtained by a Monte Carlo simulation.
The red, green, and blue points respectively indicate $\boldsymbol{\eta}={\bf e}_1,{\bf e}_2,{\bf e}_3$, and the white points correspond to $\boldsymbol{\eta}=\boldsymbol{0}$.
Above the transition temperature, many white points appear and the colored points are positioned randomly.
As the temperature decreases, the number of white points decreases.
Below the transition temperature, red points spread over the system, which means threefold symmetry breaks at the transition point.
These behaviors of snapshots resemble those observed in the 3-state Potts model with invisible states shown in Ref.~\cite{Tanaka-2011a}.

We also consider the relation between the model given by Eq.~(\ref{eq:Hamiltonian}) and the $q$-state Potts model with invisible states, where the Hamiltonian is given by
\begin{eqnarray}
{\cal H}_{\rm Potts} = -J \sum_{\langle i,j \rangle} \delta_{\sigma_i,\sigma_j}
 \sum_{\alpha=1}^q \delta_{\sigma_i,\alpha},
 \quad
 \sigma_i = 1, \cdots, q, q+1, \cdots, q+r,
\end{eqnarray}
where $r$ is the number of invisible states.
Since the long-ranged correlation can be ignored at high temperatures, it is enough to consider a two-spin system.
In the two-spin system, correlation between two spins $\langle \delta_{\sigma_1,\sigma_2}\rangle$ is expressed as
\begin{eqnarray}
 \label{eq:analysis}
\ln \left( \frac{\langle \delta_{\sigma_1,\sigma_2}\rangle}{1-\langle \delta_{\sigma_1,\sigma_2}\rangle}\right) = \frac{J}{T} - \ln (q+2r+\frac{r^2}{q}-1).
\end{eqnarray}
To consider the relation in the model defined by Eq.~(\ref{eq:Hamiltonian}), we calculate the correlation of local parameters between nearest-neighbor triangle pairs.
Figure \ref{fig:snapshot} (c) shows the LHS of Eq.~(\ref{eq:analysis}) with $q=3$ as a function of $1/T$ for several $\Xi$'s in the high-temperature region.
The lines are obtained by the least-squares estimation.
The good fits in Fig.~\ref{fig:snapshot} (c) mean the model given by Eq.~(\ref{eq:Hamiltonian}) in the high-temperature region is described well by the 3-state Potts model with invisible states using our method.

Next we consider the case that the order parameter space is SO(3)$\times Z_2$, which has a second-order phase transition with twofold symmetry breaking.
In this case, there are two types of ground states ($\alpha=1,2$), as shown in Fig.~\ref{fig:snapshot} (d).
Since we are considering the nature of a phase transition with twofold symmetry breaking, a local parameter should be defined so as to be similar to the 2-state Potts model, {\it i.e.} the Ising model.
Using $\xi^\alpha$, a local (scalar) parameter $\eta$ can be defined by $\eta=(-1)^{\alpha'+1}H(\xi^{\alpha'}-\Xi)$.
Then the phase transition with twofold symmetry breaking can be described via the $2$-state Potts model with invisible states.
We use the parameter sets $J_1/J_3=-0.4926$ and $\lambda =1.308$, in which the transition temperature is $T_{\rm c}/J_3=0.4950(5)$ and the relative angles $\{\theta_\ell^{\alpha}\}$ in the ground state are shown in Fig.~\ref{fig:snapshot} (d)\cite{Tamura-2013}.
Figure~\ref{fig:snapshot} (e) shows snapshots of the local parameters in the case of $\Xi=0.8$ at several temperatures.
The black and yellow points respectively indicate $\eta=1,-1$ whereas the white points indicate $\eta=0$.
As well as the previous case, the number of white points decreases as the temperature decreases.
Below the transition temperature, the black points spread over the system, which means twofold symmetry breaks at the transition point.
Figure~\ref{fig:snapshot} (f) shows the LHS of Eq.~(\ref{eq:analysis}) with $q=2$ as a function of $1/T$ for several $\Xi$'s in the high-temperature region.
The lines are obtained by the least-squares estimation.
For a certain parameter region $\Xi$, there is a good fit as in the case that the order parameter space is SO(3)$\times C_3$.

In this paper, we demonstrated a microscopic analysis of the phase transitions in the $J_1$-$J_3$ model on a distorted triangular lattice. 
We believe that our proposed method is useful in analyzing a phase transition with discrete symmetry breaking in frustrated continuous spin systems.

We thank Hikaru Kawamura, Michikazu Kobayashi, Seiji Miyashita, Satoru Nakatsuji, Takafumi Suzuki, Yusuke Tomita, and Hirokazu Tsunetsugu for useful comments and discussions.
The authors are partially supported by National Institute for Materials Science (NIMS), Grand-in-Aid for Scientific Research (C) (25420698), JSPS Fellows (23-7601), Scientific Research (B) (22340111), and the Computational Materials Science Initiative (CMSI). 
Numerical calculations were performed on supercomputers at the Institute for Solid State Physics, University of Tokyo.

\end{document}